# Broadband Tunable Terahertz Polarization Converter Based on Graphene-Shaped Metasurfaces

Behnaz Bakhtiari
*School of Electrical Engineering*
*Iran University of Science and Technology*
Tehran, IRAN
b_bakhtiari@elec.iust.ac.ir

Homayoon Oraizi
*School of Electrical Engineering*
*Iran University of Science and Technology*
Tehran, IRAN
h_oraizi@iust.ac.ir

Ali Abdolali
*School of Electrical Engineering*
*Iran University of Science and Technology*
Tehran, IRAN
abdolali@iust.ac.ir

*Abstract*—In this paper, a broadband tunable polarization converter based on graphene metasurfaces is proposed. This polarization converter works in the terahertz (THz) frequency region, using the advantage of graphene characteristics to have a tunable frequency response. The designed graphene-shaped periodic structure on top of the substrate is utilized to convert the incident wave polarization to the desired target in a flexible operational band in the THz frequencies. The polarization conversion ratio is more than 0.85 in a wide range of frequencies in the THz band from 4.86 to 8.42 THz (the fractional bandwidth is 54%). The proposed polarization converter is insensitive to the angle of the incident wave up to 40°. Using graphene provides a tunable frequency response without changing the geometry of the designed structure.

## I. INTRODUCTION

In recent years there is an increasing interest in attempts to design and develop devices in the THz frequency range, which is between microwave and infrared frequency regions. THz technology and devices can be applied in sensing, detecting, and many telecommunication, medical, and optical applications.

Control or manipulation of the electromagnetic wave to the desired target is one of the fundamental approaches to design telecommunication devices [1]. Polarization, as one of the characteristics of the electromagnetic wave, can be assumed for manipulating electromagnetic waves to the desired goal. Polarization converters are widely used to convert polarization of electromagnetic or optical waves to the desired target; consequently, there are many applications in sensing, detecting, imaging, and other communication and optical applications [2-3]. The polarization converter in the THz region can be a demand to make this region applicable as the growing interest to develop microwave devices to function in the THz frequencies.

Recently, metamaterials that are bulky structures to manipulate electromagnetic waves are widely employed to develop devices, providing extraordinary properties that cannot be found in natural materials [4]. However, the complexity of the design, loss, and massive structure of these devices lead scientists to manipulate electromagnetic waves using metasurface structure, which is a two-dimensional periodic structure, to control and manipulate electromagnetic waves [5-6].

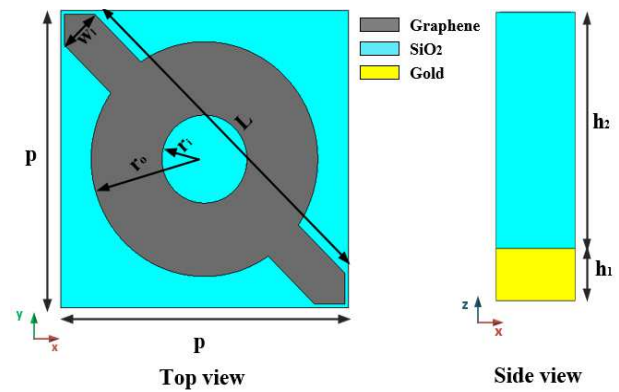

Fig. 1. Unit cell and geometry of the proposed polarization converter.

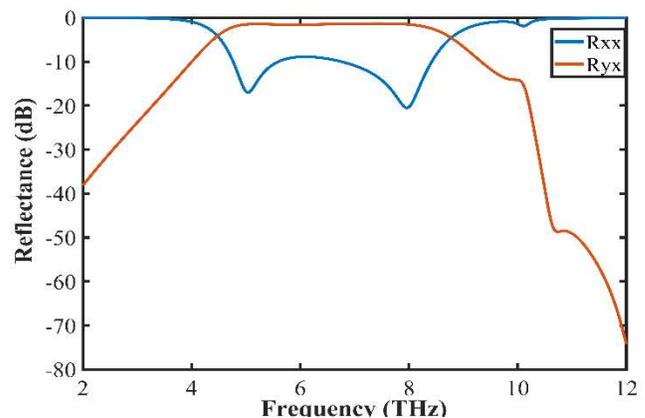

Fig. 2. The frequency response of the reflectances

Graphene is a 2D-one-layer material consists of the honeycomb structure of atoms, which can be used to optimize and develop the function of devices because of its unusual electrical characteristics.

Graphene, as an advanced electromagnetic material, can be used in electromagnetic devices to enhance and develop their functionality [7].

There are two methods to increase the frequency response bandwidth of polarization converter devices, using geometrical structures with multiple resonances [5-7] or using multiple layer structures [8].

In this paper, a one-layer polarization converter device based on metasurface structure is proposed, which gains the usage of graphene to have a tunable polarization conversion work in the THz frequencies region.

## II. DESIGN AND SIMULATION

In this paper, a polarization converter metasurface is designed to convert the polarization of incident waves in a wide range of frequency in the THz band besides having an adjustable frequency response using graphene's benefit characteristics in the designed structure.

For the design process, the reflectance frequency response is chosen to use the advantage of the single-layer structure. The reflection coefficient of the polarization converter is defined by its scattering matrices (1):

$$R = \begin{bmatrix} R_{xx} & R_{xy} \\ R_{yx} & R_{yy} \end{bmatrix} \quad (1)$$

As the proposed polarization converter is symmetric consequently, $R_{xx}=R_{yy}$ and $R_{xy}=R_{yx}$. $R_{xy}$ is the reflection coefficient when the reflected electric field is along the x-direction and the incident electric field is along the y-direction.

The polarization conversion ratio, as demonstrated in (2), is a function of co-polarization ($R_{xx}$) and cross-polarization ($R_{xy}$). As illustrated in (2), it is necessary to reduce the value of co-polarization reflectance to reach the best result for the value of polarization conversion. Many different attempts are needed to find key geometrical parameters that affect the result to reach optimal frequency response to provide a wideband polarization converter with a perfect performance.

$$PCR = \frac{|R_{yx}|^2}{|R_{xx}|^2 + |R_{yx}|^2} \quad (2)$$

The result of the polarization conversion ratio from the structure is monitored using CST Microwave Studio software, which is a software for the design and analysis of 3D electromagnetic structures. An optimal design structure can be determined by analyzing the proposed structure in the eigenmode solver of CST Microwave Studio and localizing the plasmon resonances of the structure by altering values of parameters from the structure's geometry. As shown in Fig. 1 a broadband polarization converter based on metasurface is designed. The proposed polarization converter consists of a periodic structure, including the replications of the depicted unit cell. A graphene shaped structure is located on the top of the SiO$_2$ substrate with $\varepsilon_r$ =2.1, μ=1, and thickness of 8.5 um (d$_2$).

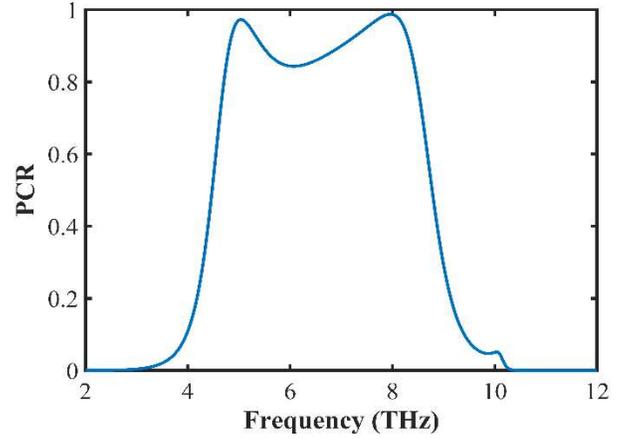

Fig. 3. The polarization conversion ratio of the proposed polarization converter

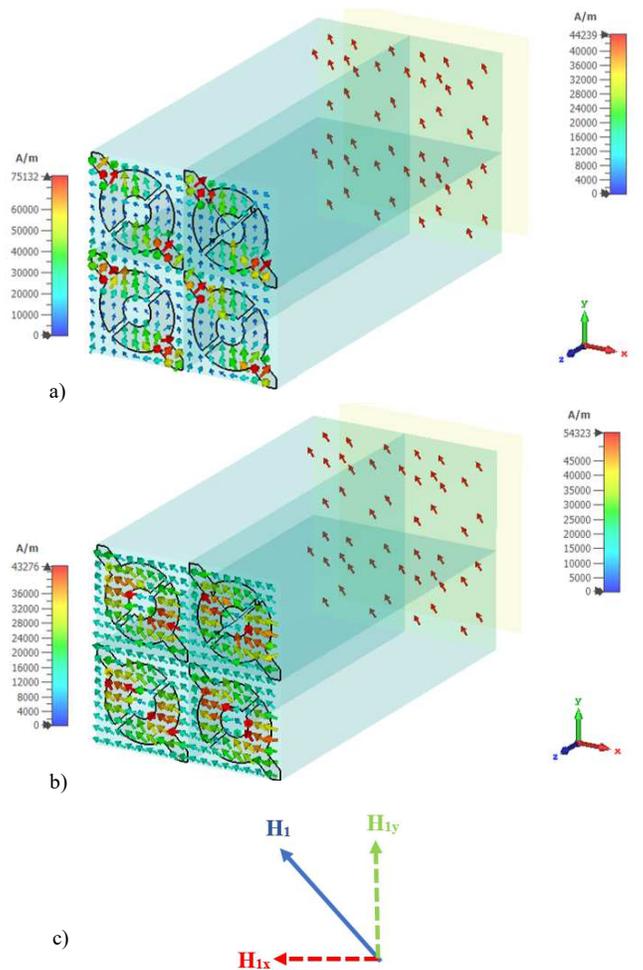

Fig. 4. Surface currents on top and the bottom layer at surface plasmon resonance at a) 5.05 THz b) 7.95 THz c) the induced magnetic field

A Gold sheet with σ= 4.56×10⁷ S/m and thickness of 1 um ($d_1$) on the back of the structure exists. The period of the unit cell is 2 um (p), and other values of parameters are listed below:

l=2.5um, $r_i$=0.3um, $r_o$=0.8um, $w_l$=0.3um, and $w_g$=0.1um

Graphene chemical potential (μ) =0.6 eV

Graphene relaxation time (τ) =0.6 ps

Graphene temperature (T) =300 k

In Fig. 2 frequency response of co-polarization and cross-polarization is shown and PCR reaches the optimal results which have the frequency range from 4.86 to 8.42 THz, with a polarization conversion ratio of more than 85% in a wide range of frequency in the THz frequency band (Fig. 3). There are two resonance modes in 5.05 and 7.95 THz.

Fig. 4 shows the surface currents on top and bottom layers of the proposed polarization converter, in which these surface currents at 5.05 THz are in the anti-parallel direction, generating the magnetic resonance (Fig. 4a) and in the parallel direction, enhancing the electric resonance at 7.95 THz (Fig. 4b). When the incident electric field is along the x-direction, surface currents direction at 5.05 THz leads to an induced magnetic field along with $H_1$, which can be decomposed to two orthogonal vectors (Fig. 4c). since $H_{1y}$ is along $H_y$ it cannot generate any cross-polarization but, $H_{1x}$ can convert polarization of the incident electric field to $E_y$ direction, consequently, the reflected wave is y-polarized.

*A. Tunability by graphene*

Graphene is a two-dimensional material with various electrical, chemical, and mechanical characteristics that have made it an excellent candidate to use in engineering.

The surface conductivity of graphene is a function of different parameters that can be tuned by their variations. As demonstrated in the Kubo formula [9], the conductivity of graphene can be controlled by changing the value of these parameters by electrical or magnetic biasing or chemical doping.

As demonstrated in (3), the conductivity of graphene consists of two parts called interband and intraband conductivity.

$$\sigma_s = \sigma_{intra-band}(\omega,\mu_c,\Gamma,T) + \sigma_{inter-band}(\omega,\mu_c,\Gamma,T) \quad (3)$$

$$\sigma_{intra-band}(\omega,\mu_c,\Gamma,T) = \frac{-je^2 k_B T}{\pi\hbar^2(\omega-j2\Gamma)}(\frac{\mu_c}{k_B T} + 2\ln(e^{-\mu_c/k_B T}+1))$$

$$\sigma_{inter-band}(\omega,\mu_c,\Gamma,T) = -j\frac{e^2}{4\pi\hbar}\ln(\frac{2|\mu_c|-(\omega-j2\Gamma)\hbar}{2|\mu_c|+(\omega-j2\Gamma)\hbar})$$

In (3) e is the electron charge, $k_B$ is Boltzmann constant, h is reduced Plank constant, T is temperature, $\mu_c$ is Fermi energy or chemical potential, $\Gamma=\tau^{-1}$, and τ is relaxation time.

*B. optimization*

In this paper period of repetition and thickness of the substrate were chosen as key parameters for the design process. So, by altering the value of the period from 2 to 2.3 um, which is shown in Fig. 5, and changing the value of substrate thickness from 7 to 8.5 um (Fig.6.) the dimensions for the optimal frequency response is reached (as explained before).

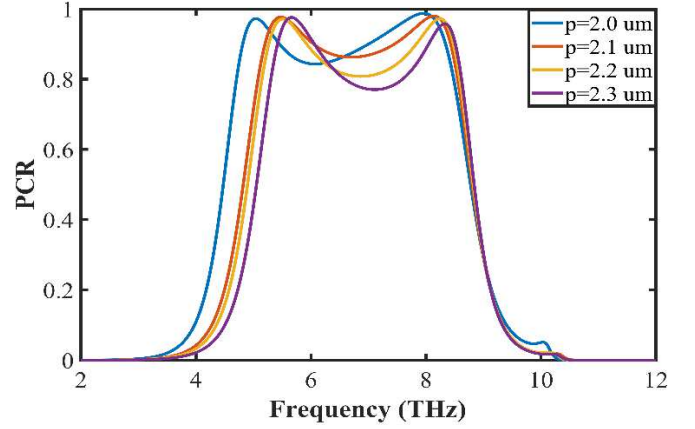

Fig. 5. The polarization conversion ratio for different values of the unit cell's period

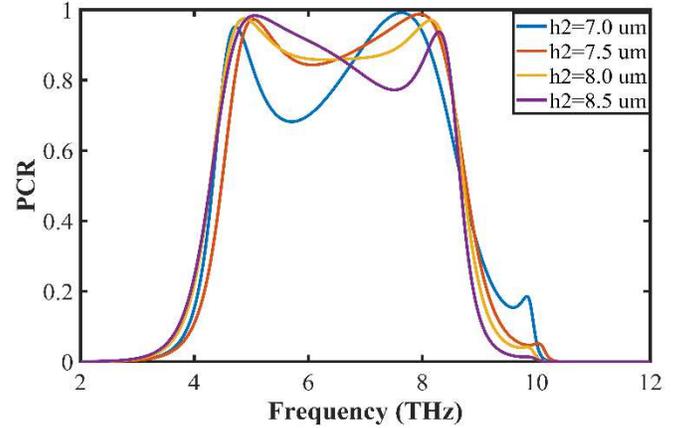

Fig. 6. The polarization conversion ratio for different value of substrate thicknesses

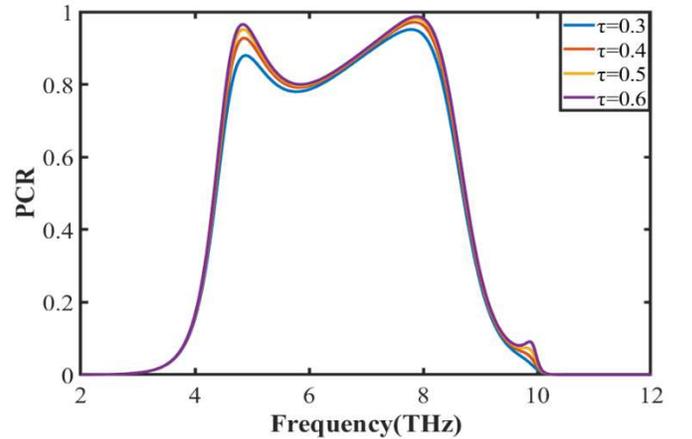

Fig. 7. Tunable polarization converter controlled by the relaxation time of graphene

## III. RESULTS

Using a graphene-shaped pattern in the structure of metasurfaces provides the possibility of controlling frequency response by the graphene's parameters. As shown in Fig. 7. various values of polarization conversion can be reachable by altering the relaxation time of graphene. The polarization conversion ratio can be controlled by the different chemical potential of graphene (Fig. 8), in other words, by applying electrical bias, the frequency band of polarization conversion can sweep different frequency region in THz band, which provide the ability to convert polarization in a different range of frequencies with just one geometry and without changing the shape of the structure.

In Fig. 9 stability of the polarization converter with the angle of the incident wave (till 40 degrees) is depicted, which means the proposed polarization converter is insensitive to the incident angle.

## IV. CONCLUSION

This proposed polarization converter has broadband and perfect frequency response for polarization conversion with more than 90% from 5.43 to 7.66 THz besides providing a tunable frequency response in the THz band. In this paper, polarization conversion can be controlled by two approaches of altering the value of different parameters of proposed geometry, and the efficient one using a unique geometrical structure and control polarization conversion ratio by the aid of graphene characteristics. The exciting capability of tunability in polarization conversion besides a wide range of frequency response and its insensitivity to the incident angle made it an excellent candidate for a wide range of applications in sensing, detecting, and various telecommunication systems.

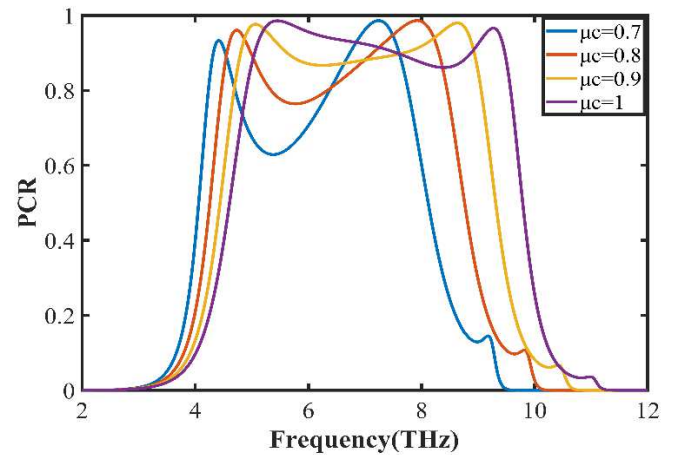

Fig. 8. Tunable polarization converter controlled by the chemical potential of graphene

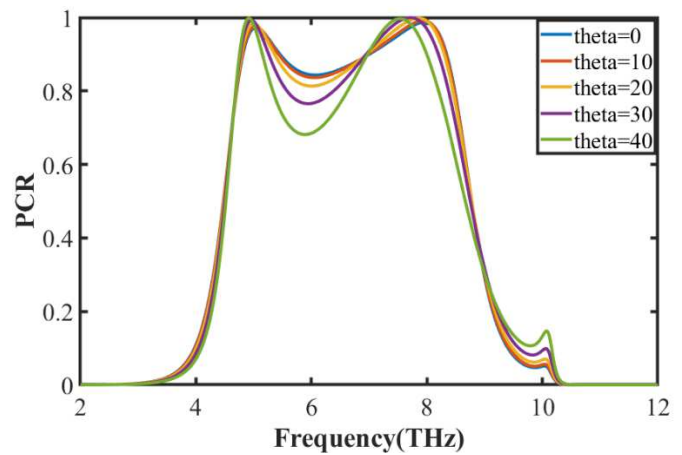

Fig. 9. Stability of polarization converter with incident angle